\documentclass[prb,aps,showpacs,twocolumn]{revtex4}
\usepackage{graphicx,amsfonts,bm,amsmath}
\newcommand{\rl}{\rangle\!\langle}
\DeclareMathOperator{\tr}{Tr}

\begin{document}
\title{Manifestation of fundamental quantum complementarities 
in time-domain interference experiments with quantum dots: a
theoretical analysis}
\author{Pawe{\l} Machnikowski}
 \email{Pawel.Machnikowski@pwr.wroc.pl}
 \affiliation{Institute of Physics, Wroc{\l}aw University of
Technology, 50-370 Wroc{\l}aw, Poland}

\begin{abstract}
A theoretical analysis is presented showing that fundamental
complementarity between the particle-like properties of an exciton
confined in a semiconductor quantum dot and the ability of the same
system to show interference may be studied in a time domain
interference experiment, similar to those currently performed. The
feasibility of such an experiment, including required pulse parameters
and the dephasing effect of the environment, is studied.
\end{abstract}


\maketitle

\section{Introduction}

Bohr's complementarity principle \cite{bohr28} is one of the 
central points of the
quantum theory. It states that laws of nature allow one to describe
a quantum system in terms of mutually excluding properties. Such
apparently contradictory, complementary descriptions are indispensable
for complete understanding of the quantum world but the
corresponding properties of a physical system can 
never fully manifest themselves simultaneously in a single experiment.

The most celebrated complementarity of this kind is the particle--wave
duality. The usual starting point for many textbooks is to discuss it
in terms of a double-slit (Young) experiment, in which particles form
an interference pattern on a screen after passing through a system of two
openings. This wave-like behavior disappears if one detects
through which of the openings the particle \textit{really} passed,
thus querying a particle-like property (indivisibility). In fact, 
visibility reduction results from correlations arising between
the particle and its environment \cite{scully91}, 
due to which a measurement on the
latter might, in principle, yield information on the path chosen by
the particle (\textit{which path} information).
Thus, interference is affected no matter whether this measurement is
actually done and the effect is the same for controlled 
measurement (environment being part of a measurement apparatus) and
for uncontrolled environment-induced dephasing. 

Experimentally pursuing the relation between the available
\textit{which path} 
information and the ability to show interference presents a twofold
difficulty. First, one needs sufficient quantum control over an
individual quantum system to observe quantum interference
effects in a single-particle experiment. 
Second, in order to study the whole range of intermediate
cases, where some partial information on the particle path is
attained, one has to be able to extract information on the quantum
state (path) of the system not only in a non-destructive manner but
also in a controlled amount, i.e., in a way that allows one to infer
the path correctly only with a certain probability $1/2<p<1$. Ideally,
the extracted information should be experimentally available (which is
not the case if the systems is ``measured'' by the environment in a
dephasing process) so that its amount may be independently verified.

In spite of these challenging requirements, the effect of partial
\textit{which path} information on visibility of interference fringes
was observed in optical \cite{zou91,brune96}, neutron
\cite{summhammer87} and atomic
\cite{scully91,pfau94,chapman95,durr98} interference
experiments. \textit{Which path} decoherence of electrons has been tested in
coherent transport experiments in semiconductor systems
\cite{aleiner97,buks98a,avinun04}. An experiment with a free electron
interferometer \cite{hasselbach88} has also been proposed 
\cite{anglin97,hasselbach00}.

It is natural to investigate quantum complementarity 
by studying \textit{which path} decoherence in
space domain experiments, where the two system states correspond to
real trajectories (paths) of a particle. However,  
 phase-sensitive superpositions and the
resulting interference effects are manifested also in 
time-domain interference experiments, like
those\cite{bonadeo98,kamada01,htoon02} performed on excitons 
(electron-hole pairs) confined in 
semiconductor quantum dots (QDs) \cite{jacak98a,bimberg99}. 
Here, the final exciton 
occupation (probability of
finding the exciton in the QD over a large number of repetitions) 
after a sequence of two
optical pulses oscillates as a function of the relative phase 
(time delay) between the pulses. Obviously, in spite of the
wave-like behavior manifested by this interference effect, 
a single occupation measurement always yields either 0 or
1, demonstrating the particle-like nature of the exciton.
In view of the important role that quantum complementarity plays in
the fundamentals of quantum mechanics it seems interesting to
take advantage of the recent progress in time domain interference
experiments on QDs and to design an experiment in which quantum
complementarity manifests itself in these systems.

In this paper it is shown that the existing time-domain interference
experiments on QD systems may indeed be extended to test complementarity
between the knowledge on the particle-like state of an exciton  and 
its ability to show quantum interference. To
this end, one performs a detection of the exciton between the two
pulses, when the system is in a superposition state. This is
equivalent to determining through which slit a particle passes in a
Young setup and corresponds to testing a particle-like property of the
confined exciton (is the exciton \textit{really} there?) treating it
as an indivisible entity. Similarly as in the space-domain setup, this
must affect the ability of the same system to show wave-like behavior
(interference), in accordance with the fundamental complementarity
principle. A non-destructive method of extracting partial information
on the system state is provided in a natural way by a non-projective
indirect measurement, via conditional dynamics of a biexciton system
leading to a variable degree of correlation with a second
exciton (with opposite polarization or confined in a neigboring dot
with different transition energy) that plays the role of an ancilla
system.  Depending on the amount of extracted knowledge on the system
state, the visibility of interference fringes is reduced to a various
degree, so that the complementarity principle may be studied on the
quantitative level.  Moreover, the extracted information is accessible
via an additional measurement (up to imperfect, but known, detector
efficiency) and its amount may be verified so that experimental access
to both components of the complementarity principle is provided. In
the paper, an appropriate sequence of control pulses for performing
the complementarity experiment is discussed. The feasibility of the
experiment in terms of spectral selectivity of control pulses and of
resilience of the effect against environmental dephasing is also
studied.

The paper is organized as follows. In Sec.~\ref{sec:interf} time
domain interference experiments on QDs are reviewed. 
Sec.~\ref{sec:wp}
describes the experiment in which the complementarity may be
tested. The feasibility of the coherent control necessary for
realizing the proposed scheme under realistic conditions
is studied in Secs.~\ref{sec:feasib} and
\ref{sec:decoh}. The 
final section contains concluding remarks, including a brief discussion of
available experimental techniques.

\section{Time domain interference}
\label{sec:interf}

Let us start with a brief review of time domain interference
experiments on quantum dots. QDs are artificial atomic-like
nanostructures with carrier states quantized due to strong spatial
confinement \cite{jacak98a,bimberg99}. The fundamental optical
transition in a QD consists in promoting an electron from the valence
band to the conduction band, thus creating a confined electron-hole
pair, referred to as a (\textit{confined}) \textit{exciton},
corresponding to a single spectral line. In an ideally circular dot,
the confined exciton state is
degenerate with respect to the orientation of the spins of the
carriers. A single transition out of the degenerate doublet may be
addressed by an appropriate choice of circular polarization of the laser beam.
In a more general case, breaking of the circular symmetry leads to
mixing of the angular momentum eigenstates due to electron-hole
exchange interaction \cite{bayer02b} and to a
fine structure splitting. In the presence of this effect, excitons
with definite circular polarization are no longer eigenstates of the
system and an oscillation between the two polarization states will
appear. Nonetheless, since the fine structure splitting is usually
very small ($\sim 100\;\mu$eV or less), the period of these
oscillations is relatively long and its effect on the system dynamics
may be neglected for sufficiently short pulse sequences.

Time domain interference experiments on 
QDs\cite{bonadeo98,kamada01,htoon02} are performed with two
phase-locked laser pulses selectively tuned to one of the two 
fundamental optical transitions. The first pulse induces a
superposition of no exciton and single exciton states. The second
pulse rotates the system state further, with a phase shift depending
on the delay between the pulses. The final occupation 
varies periodically as a function of the phase shift, 
producing time-domain interference fringes.

We will denote the empty dot state by $|0\rangle$ and the single
exciton state by $|1\rangle$. The exciton energy is
$\epsilon_{\mathrm{S}}$ (`S' stands for `system' and refers to the
exciton transition addressed in the interference experiment). 
In the rotating basis, 
$|\tilde{1}\rangle=e^{i\epsilon_{\mathrm{S}}t}|1\rangle$,
$|\tilde{0}\rangle=|0\rangle$,
the Hamiltonian for the driven system, upon neglecting the non-resonant
terms (rotating wave approximation, RWA), is
\begin{eqnarray}\label{ham1}
H_{1} & = &  
\frac{1}{2}\mu E_{\mathrm{S1}}(t)
(|\tilde{0}\rl \tilde{1}|+\mathrm{H.c.})\\
\nonumber
&& +\frac{1}{2}\mu E_{\mathrm{S2}}(t-t_{\mathrm{S2}})
  (e^{-i\phi}|\tilde{0}\rl \tilde{1}| +\mathrm{H.c.}),
\end{eqnarray}
where $\mu$ is the inter-band dipole
moment, $E_{\mathrm{S1,S2}}(t)$ are the envelopes of the two
pulses, and $\phi=\epsilon_{\mathrm{S}}t_{\mathrm{S}}/\hbar$ is the
phase shift dependent on the time delay between the pulses.
The two terms in Eq.~(\ref{ham1}) account for the action of the control
pulses coupled to a single exciton transition by polarization selectivity: 
The first pulse (S1) arrives at $t=0$ and prepares the initial superposition
state. It corresponds to splitting the particle path 
in a usual space-domain (two-slit) experiment. 
The second (S2) pulse arrives at $t=t_{\mathrm{S2}}$. Its arrival time 
must be tuned with 
femtosecond accuracy to provide controllable phase-locking of the
pulses, as done in QD interference experiments
\cite{bonadeo98,kamada01,htoon02}.
This pulse plays the role of ``beam merger''
providing, at the same time, a phase shift between the ``paths''.

In the experiment, 
the system is initially in the state $|0\rangle$.
The first pulse (S1) is a $\pi/2$ pulse that performs the
transformation 
\begin{equation}\label{us1}
U_{\mathrm{S1}}=\frac{1}{\sqrt{2}}\left( 
\mathbb{I}-i\sigma_{x} \right),
\end{equation}
where $\sigma_{x}=|\tilde{0}\rl \tilde{1}|+|\tilde{1}\rl \tilde{0}|$
is the Pauli matrix.
This pulse leaves the system in the equal superposition state 
\begin{equation}\label{psi}
|\psi\rangle=\frac{|\tilde{0}\rangle-i|\tilde{1}\rangle}{\sqrt{2}}. 
\end{equation}
The second pulse is again a $\pi/2$ pulse,
\begin{equation}\label{pulse-s2}
U_{\mathrm{S2}}=\frac{1}{\sqrt{2}}\left(  
\mathbb{I}-i\hat{\bm{n}}\cdot\bm{\sigma} 
\right),
\end{equation}
where $\hat{\bm{n}}=[\cos\phi,\sin\phi,0]$ and $\bm{\sigma}$ is the
vector of Pauli matrices in the basis
$|\tilde{0}\rangle,|\tilde{1}\rangle$. 
After this pulse, the average number of excitons in the dot is 
\begin{displaymath}
N(\phi)=|\langle 1|U_{\mathrm{S2}}|\psi\rangle|^{2}
=\frac{1}{2}\left(  1-\cos\phi \right),
\end{displaymath}
and changes periodically between $N_{\mathrm{min}}=0$ and
$N_{\mathrm{max}}=1$ as a function of phase shift (or delay time),
thus producing an interference pattern.

The quality of the interference pattern is customarily quantified in terms of
visibility of interference fringes
\begin{displaymath}
{\cal V}=\frac{N_{\mathrm{max}}-N_{\mathrm{min}}}
{N_{\mathrm{max}}+N_{\mathrm{min}}}.
\end{displaymath}
The amplitude
(visibility) of the fringes in an ideal experiment is ${\cal V}=1$
for $\pi/2$ pulses, i.e., for an equal superposition in between
them. Otherwise, some \textit{a
priori} information on the superposition state can be inferred and the
visibility is reduced. For simplicity, the present discussion is
restricted to the equal superposition case.

\section{Complementarity between \textit{which path} information 
and interference}
\label{sec:wp}

This Section discusses the essential modification to time domain
interference experiments that allows one to attain
partial information on the state of the system and to observe the
related visibility reduction of the interference pattern. 
First, however, a measure of the partial information is introduced and
the complementarity principle is formulated in a quantitative 
form\cite{jaeger95,englert96}. 

The notion of ``partial information'' is understood as follows. The
system (S) of interest is coupled to another \textit{quantum
probe} (QP) system and conditional dynamics of the latter is
induced, leading to correlations between the states of the systems S
and QP. Next, a measurement on QP is performed and its result is used
to infer the state of S, i.e., to predict the outcome of a subsequent
measurement on S. The probability of a correct prediction ranges from
1/2 (guessing at random in absence of any correlations) to 1 (knowing
for sure, when the systems are maximally entangled). 
From a formal point of view, this procedure is
a non-projective, generalized measurement on the exciton
system, performed within an indirect measurement
scheme\cite{breuer02}.

Quantitatively, an intrinsic measure of information on
the system S extracted by QP is provided by the
\textit{distinguishability of states}\cite{jaeger95,englert96},
\begin{equation}\label{disting-def}
{\cal D}=2\left(p-\frac{1}{2}\right), 
\end{equation}
where $p$ is the probability a correct prediction for the state of S
maximized over all possible measurements on QP.
In this way, guessing at random and knowing for sure correspond to 
${\cal D}=0$ and ${\cal D}=1$, respectively. 
According to a general theory \cite{jaeger95,englert96},
the complementarity relation between the 
knowledge of the system state and the visibility of the fringes may be
written, using the distinguishability ${\cal D}$ as a measure of
information, in the quantitative form, 
\begin{equation}\label{ineq}
{\cal D}^{2}+{\cal V}^{2}\le 1.
\end{equation}
The equality holds for systems in pure states. 

In a QD, the formal scheme of indirect measurement translates
naturally into well known conditional dynamics of a biexcitonic system
in which the exciton addressed in the interference experiment
described in Sec.~\ref{sec:interf}  (system S) is coupled to another
exciton (QP), localized either in the same or in a neighboring dot. In
the former case the subsystems are distinguished by their
polarization, while in the latter case they are distinguishable by
different excitation energies.
The Coulomb (dipole-dipole) interaction between the two
excitons shifts the energy of the biexciton
state\cite{hartmann00a}, so that spectrally narrow pulses
may induce dynamics of the QP exciton conditional on the state of the
other one, as required for the indirect measurement scheme 
\cite{li03,unold05}.
On the other hand, spectrally broad pulses
may be used to perform unconditional rotations. The final measurement
is done by detecting photons emitted by recombining excitons, again
with polarization or energy resolution. While it is reasonable to
neglect the effect of the fine-structure splitting over the relatively
short duration of the control sequence, 
the average time before the photons are emitted (exciton lifetime) is
usually longer (hundreds of picoseconds to nanoseconds) and
the non-conservation
of the exciton polarization may
have considerable impact on measurement results in the single-dot
scheme based on polarization-selectivity (this restriction 
may be partly overcome, as explained below).

\begin{figure}[tb]
\begin{center}
\unitlength 1mm
\begin{picture}(85,45)(0,5)
\put(0,0){\resizebox{34mm}{!}{\includegraphics{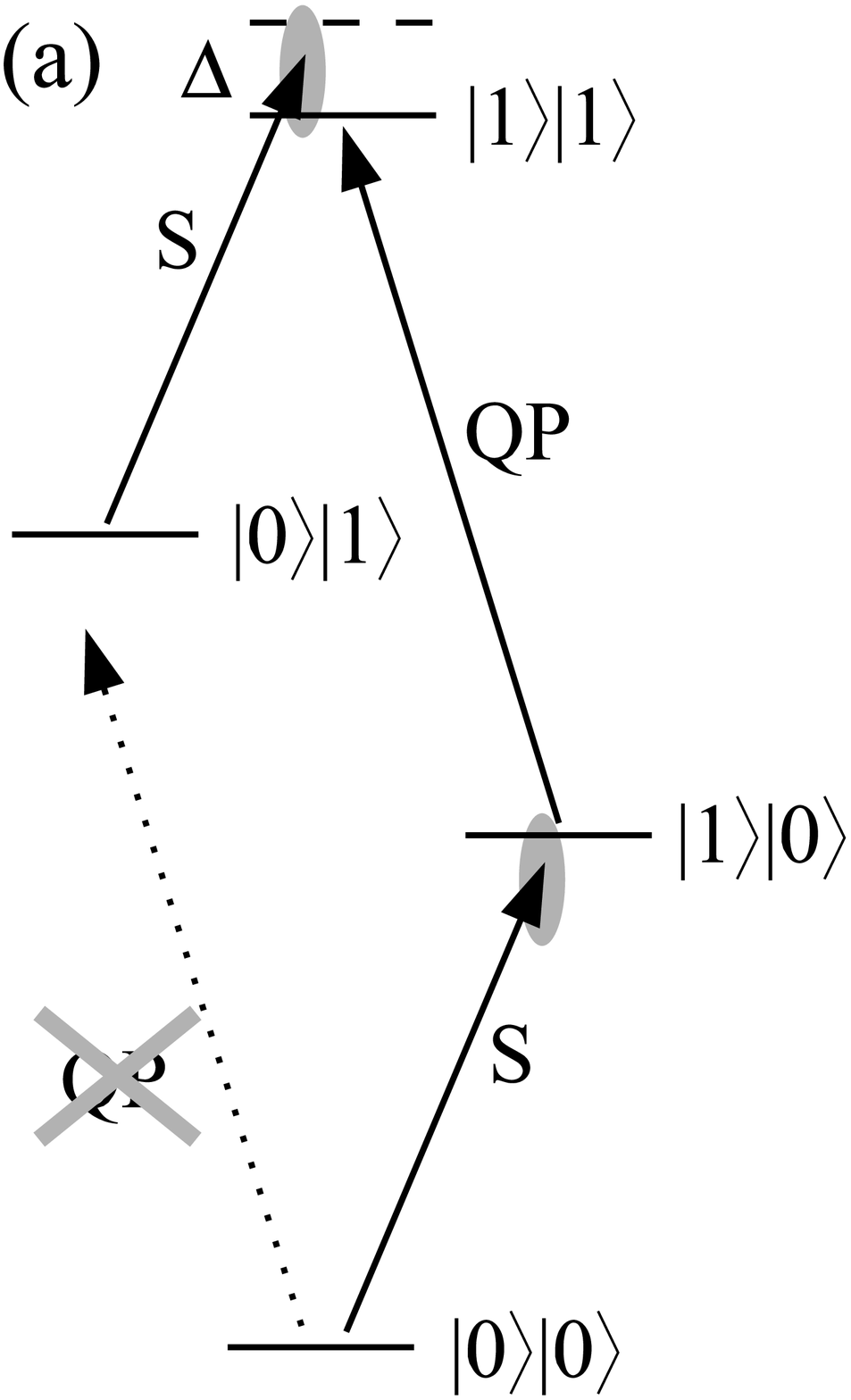}}}
\put(34,0){\resizebox{51mm}{!}{\includegraphics{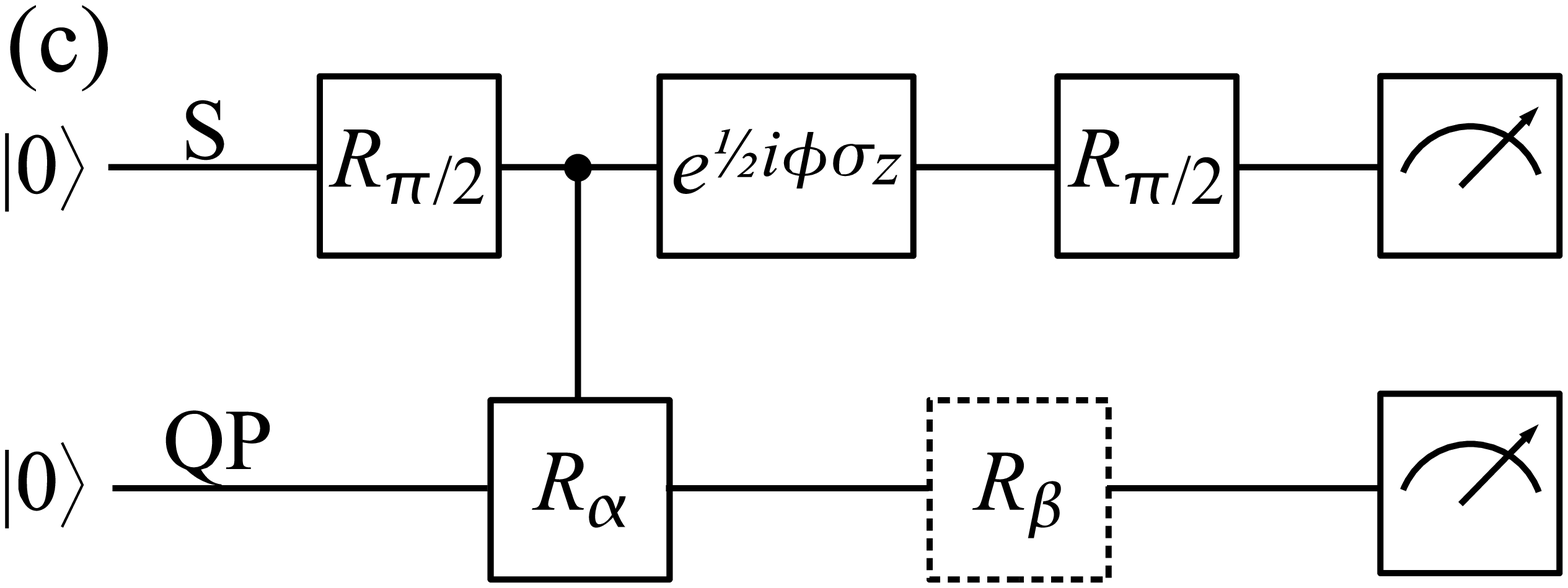}}}
\put(34,25){\resizebox{51mm}{!}{\includegraphics{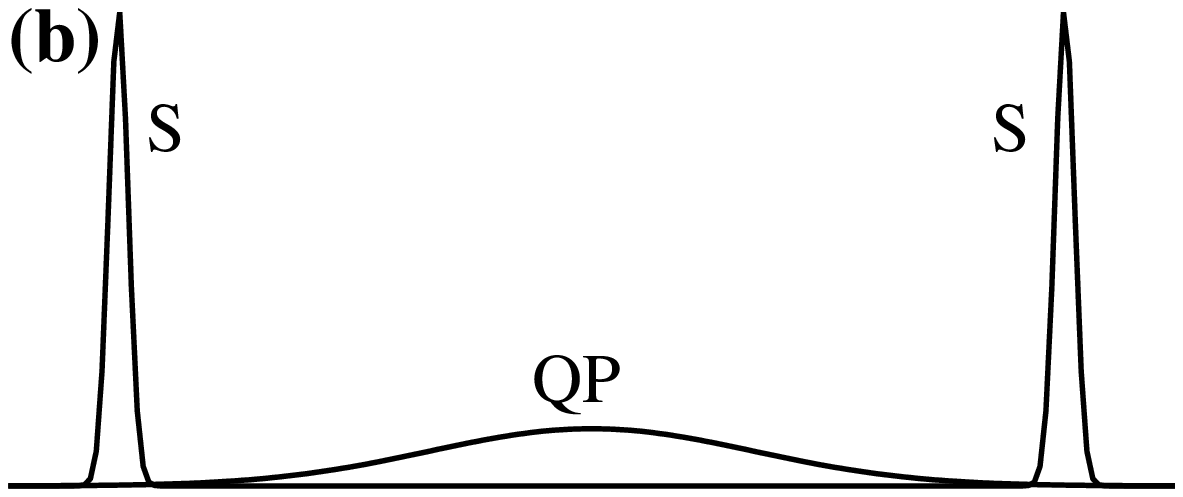}}}
\end{picture}
\end{center}
\caption{\label{fig:diagrams}(a) The diagram of energy levels and
transitions in the system, assuming that the excitons are confined in
neighboring dots with different transition energies. The
pulses inducing transitions on the system S have broad spectrum so
that both transitions are possible. The pulse acting on
the QP system is spectrally selective and tuned to the biexcitonic
transition, so that the single-exciton transition is
energetically forbidden in this subsystem. 
(b) The sequence of pulses used in the experiment.
(c) Quantum-logic diagram \cite{nielsen00} of
the performed operations. Here 
$R_{\alpha}=\cos(\alpha/2)\mathbb{I}-i\sin(\alpha/2)\sigma_{x}$.
The $R_{\beta}$ rotation on the QP subsystem (dashed box) is applied
for optimal measurement of the extracted information in the extended scheme.}
\end{figure}

We will use a tensor product notation with $|0\rangle$ and $|1\rangle$
denoting the absence and presence of the respective exciton, as
previously, with the
interfering system (S) always to the left. In the rotating 
basis with respect to both subsystems,
the RWA Hamiltonian for the biexciton system is 
\begin{eqnarray}\label{ham0}
H & = &  H_{1}\otimes\mathbb{I}
+ \Delta |\tilde{1}\rl \tilde{1}|\otimes 
|\tilde{1}\rl \tilde{1}| \\
\nonumber
&& +\frac{1}{2}\mu E_{\mathrm{QP}}(t-t_{\mathrm{QP}})
  \mathbb{I}\otimes(|\tilde{0}\rl \tilde{1}| +\mathrm{H.c.})
\end{eqnarray}
where $\Delta$ is the bi-exciton energy shift and $E_{\mathrm{QP}}(t)$
is the envelope of the pulse coupled to the QP exciton. 
Here the first 
term contains the Hamiltonian (\ref{ham1}) and
corresponds to the pulse sequence of the interference
experiment described in the previous Section, 
the second one accounts for the bi-excitonic energy shift
and the third term describes the action of the pulse coupled to the
second (QP) exciton and spectrally tuned to the exciton-biexciton
transition. This pulse arrives at $t=t_{\mathrm{QP}}$, between the
other two pulses (that is, $0<t_{\mathrm{QP}}<t_{\mathrm{S2}}$), and
will induce the conditional dynamics necessary for the indirect
measurement. Its phase is irrelevant and will be assumed 0. The
structure of system excitations, the
sequence of pulses and the corresponding quantum-logic diagram are
shown in Fig.~\ref{fig:diagrams}.

Assume that the exciton (system S) 
is in the equal superposition state (\ref{psi}).
The probability of correctly
guessing the result of a measurement in the $|0\rangle,|1\rangle$ basis
without any additional information is obviously 1/2. 
Now, we can correlate this excitonic system with the other one (QP;
initially in the state $|0\rangle$). 
To this end, one applies a selective (spectrally narrow) pulse such that 
\begin{displaymath}
\mu\int E_{\mathrm{QP}}(t)dt=\alpha.
\end{displaymath}
It performs the conditional transformation
\begin{displaymath}
U_{\mathrm{QP}}=|\tilde{0}\rl \tilde{0}|\otimes \mathbb{I}
+|\tilde{1}\rl \tilde{1}|\otimes 
\left( \cos\frac{\alpha}{2}\mathbb{I}
-i \sin\frac{\alpha}{2}\sigma_{x} \right),
\end{displaymath}
and takes the state $|\psi\rangle$ into 
\begin{displaymath}
|\psi'\rangle=\frac{1}{\sqrt{2}}|\tilde{0}\rangle\otimes|\tilde{0}\rangle
-\frac{i}{\sqrt{2}}|\tilde{1}\rangle 
\otimes\left(\cos\frac{\alpha}{2}|\tilde{0}\rangle
-i\sin\frac{\alpha}{2}|\tilde{1}\rangle\right). 
\end{displaymath}
For $\alpha=\pi$, this pulse performs
a CNOT-like transformation on the biexcitonic system. As a result, the
total system is in the maximally entangled state 
$(|0\rangle|0\rangle-i|1\rangle|1\rangle)/2$ and
a measurement on the QP system uniquely determines the state of
the system S. Hence, due to quantum correlations between the systems,
complete information on the state of S has been extracted to QP.
On the other hand, if the
biexciton is excited with a pulse with area $\alpha<\pi$ 
the correlation between the subsystems is weaker and a measurement on
QP cannot fully determine the state of S, although the
attained information may increase the probability for correctly predicting the
result of a subsequent measurement on S. According to the discussion
above, this means that partial information on the state of S is available.

In order to find the distinguishability measure in the biexciton
scheme discussed here, we write the density matrix of the total system
corresponding to the state $|\psi'\rangle$,
\begin{equation}\label{dm-full}
\varrho=\frac{1}{2}\sum_{nm}|n\rl m|\otimes\rho_{nm},
\end{equation}
with 
\begin{eqnarray*}
\rho_{00} & = & |0\rl 0|,\\ 
\rho_{11} & = & \frac{1}{2}\left( \mathbb{I}+\cos\alpha\ \sigma_{z}
-\sin\alpha\ \sigma_{y} \right),\\ 
\rho_{01}=\rho_{10}^{\dag} 
& = & i\cos\frac{\alpha}{2}|\tilde{0}\rl \tilde{0}|
-\sin\frac{\alpha}{2}|\tilde{0}\rl \tilde{1}|,
\end{eqnarray*}
where $\sigma_{i}$ are Pauli matrices. Note that $\rho_{00}$ and
$\rho_{11}$ (but not $\rho_{01}$) are density matrices.

According to the general theory \cite{jaeger95,englert96}, 
the best chance for correctly guessing the
state of S results from the measurement of the
observable $\rho_{00}-\rho_{11}$ and the probability of the correct
prediction is then 
\begin{displaymath}
p=\frac{1}{2}+\frac{1}{4}\tr |\rho_{00}-\rho_{11}|, 
\end{displaymath}
where
$|\rho|$ is the modulus of the operator $\rho$. Using the
explicit forms of the density matrices $\rho_{00},\rho_{11}$ and the
definition (\ref{disting-def}) one finds
for the distinguishability in our case
\begin{equation}\label{disting}
{\cal D}=\frac{1}{2}\tr \left|\rho_{00}-\rho_{11}\right|
=\left|\sin\frac{\alpha}{2}\right|.
\end{equation}
Thus, the amount of information on the system S accessible via
a measurement on QP increases from 0 (no QP pulse at all) to 1 (for a
$\pi$ pulse).

Next, we study the effect of extracting the which path information 
on the interference fringes. In the state (\ref{dm-full}),
the reduced density matrix of the subsystem S is 
\begin{displaymath}
\rho_{\mathrm{S}}=\tr_{\mathrm{QP}} \varrho=
\frac{1}{2}\left( \mathbb{I}-\cos\frac{\alpha}{2}\sigma_{y} \right).
\end{displaymath}
Upon applying the second pulse of the interference experiment scheme,
namely the unconditional $\pi/2$ pulse [Eq.~(\ref{pulse-s2})],
the average number of excitons in the dot is 
\begin{displaymath}
N(\phi)=
\langle 1|U_{\mathrm{S2}}\rho_{\mathrm{S}}U_{\mathrm{S2}}^{\dag}|1\rangle
=\frac{1}{2}\left(  1-\cos\frac{\alpha}{2}\cos\phi \right).
\end{displaymath}
Now, the average occupation oscillates between the limiting values 
$(1\pm|\cos\alpha|)/2$ and the visibility of the fringes is
\begin{equation}\label{visib}
{\cal V}=|\cos(\alpha/2)|. 
\end{equation}
Comparing Eq.~(\ref{visib}) with Eq.~(\ref{disting}) it is clear that the
more certain one is whether the exciton \textit{is there} or
\textit{not} ($\cal D$ increases), the less clear the interference fringes
become ($\cal V$ decreases). 
Quantitatively, the relation ${\cal D}^{2}+{\cal V}^{2}=1$
holds which is consistent with the complementarity relation (\ref{ineq}).
As we will see below, in the presence of coupling to the environment, 
where both subsystems are in mixed states due to dephasing, this
relation will turn into inequality.

Let us notice that the impact of the which path information on
interference fringes is the same no matter
whether the QP subsystem is measured before or after generating
and detecting the interference fringes and even whether it is measured
at all. In fact, the result of any such measurement (assuming perfect
detectors) is known in advance since the state of this subsystem is
completely determined by the area $\alpha$ of the QP pulse. 
This allows the essential part
of the experiment to be performed even without polarization-selective
measurement (e.g., using the current measurement in a photo-diode
structure \cite{zrenner02}), by subtracting the known quantum probe
exciton contribution from the detected signal. The same procedure may
be applied if the exciton polarization is not conserved over the
exciton lifetime which precludes polarization-resolved optical measurement in
the single-dot setup.

With polarization-resolved detection or in a two-dot setup with
spectral resolution, it is possible to demonstrate that
information contained in the quantum probe system indeed increases the
chances of correctly guessing the state of the first excitonic system
to the extent predicted by the theory. To this end, one measures
directly both exciton occupations, without applying the third pulse
(S2), by registering both emitted photons with
time-tagging enabling the identification of coincidences. The optimal
measurement of the second exciton state should be done in the basis of
eigenstates of $\rho_{00}-\rho_{11}$, which are
\begin{displaymath}
|{\pm}\rangle=\frac{\sqrt{1\pm\sin(\alpha/2)}}{\sqrt{2}}|\tilde{0}\rangle
\mp i \frac{\sqrt{1\mp\sin(\alpha/2)}}{\sqrt{2}}|\tilde{1}\rangle.
\end{displaymath}
Since detection of photons emitted
during exciton recombination corresponds to measurement in the
$|0\rangle,|1\rangle$ basis, unconditional rotation to the optimal
basis is necessary. This is done by a pulse with area $\beta$ such that
$\cos\beta=\sqrt{1+\sin(\alpha/2)}/\sqrt{2}$ and phase exactly
opposite to that of the QP pulse. Upon detecting a photon from the
quantum probe exciton one ``predicts'' that the state of the first subsystem
should be $|1\rangle$ and vice versa. The fraction of correct
predictions is equal to the fraction of coinciding measurements
(detection or no detection on both polarizations). With a perfect
detector, it equals to 
\begin{displaymath}
p_{\mathrm{c}}=\frac{1}{2}\left(1+{\cal D}\right)
=\frac{1}{2}\left(1+\sin\frac{\alpha}{2}\right)
\end{displaymath}
which can
also be found directly from the state $|\psi'\rangle$. If the
detectors have efficiency $f$ and negligible dark and background count
rate than the actual coincidence rate may still be recovered from the
detection coincidence rate $p_{\mathrm{c}}'$ upon removal of
detection error asymmetry by preceding half of the measurements with
an unconditional $\pi$ rotation on both subsystems. A simple analysis
of the relevant conditional probabilities then yields
$p_{\mathrm{c}}=(p_{\mathrm{c}}'+f-1)/f^{2}$. 

\section{Feasibility analysis}
\label{sec:feasib}

So far, the basic scheme for testing the relation
(\ref{ineq}) in an optical interference experiment on a QD was
discussed in an idealized case, assuming perfectly
selective or non-selective pulses, as required, and neglecting
dephasing. In this Section we study the feasibility of
coherent control required for demonstration of the complementarity
principle from the point of view of spectral selectivity of the pulses. 

The short pulses must provide for unconditional excitation of the
first subsystem, so their spectral width must be larger than the
bi-excitonic shift.  Otherwise, the dynamics of the interfering system
is conditioned on the detection (QP) subsystem and the results cannot
be interpreted in terms of quantum complementarity. On the contrary,
the QP pulse must distinguish between the single exciton and
bi-exciton transition in order to act as an efficient detector. If
this distinguishability fails, the actual amount of gained information
is lower than the \textit{a priori} value of $\sin(\alpha/2)$ and the
fringe visibility remains higher than expected. Finally, the overall
duration of the pulse sequence should not be too long because of the
competition of the dephasing processes (even on 10 ps time scales in
some QD systems \cite{guenther02,chen02,li03}) and of the polarization
rotation due to electron-hole exchange interaction. 

First, we simulate the system behavior without dephasing in
order to verify that the desired spectral selectivity may be achieved
for typical system parameters. The simplest way to assure spectral
selectivity of the QP pulse is to make it long and therefore
spectrally narrow. It turns out that the degree of selectivity is affected
by the duration of this pulse but also depends periodically on the
time delay between the pulses. In the numerical simulation the
bi-excitonic shift of $\Delta=-2.0$ meV was assumed. 
For this value,
very good agreement with the idealized prediction is
found for a Gaussian pulse with full 
width at half-maximum (FWHM) of the pulse amplitude of 2.5 ps and 0.15
ps for the long (QP) and short (S) pulses, respectively, and the
delay times between the pulses $t_{\mathrm{S}}=10.3$ ps, 
$t_{\mathrm{QP}}=t_{\mathrm{S}}/2$. The central frequency of the 
short pulses is tuned
half-way in between the exciton and bi-exciton transitions.

\begin{figure}[tb]
\begin{center}
\unitlength 1mm
\begin{picture}(50,35)(0,12)
\put(0,5){\resizebox{50mm}{!}{\includegraphics{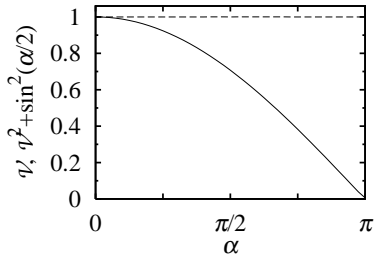}}}
\end{picture}
\end{center}
\caption{\label{fig:complem1}Solid line: 
visibility of interference fringes as a function of the QP pulse area
(determining the certainty with which the presence of the exciton is
detected).  Dashed line: the squares of visibility and
distinguishability add to 1. The plotted values are obtained from a
simulation of system dynamics, for pulse parameters as discussed in the
text, without dephasing.}
\end{figure}

The resulting visibility of the occupation
interference fringes as a function of the \textit{a priori}
distinguishability of the exciton states (determined by the area of
the QP pulse) 
is shown with a solid line in Fig. \ref{fig:complem1}. 
For all values of $\alpha$, the squares of
the visibility and distinguishability add exactly to 1 (dashed line), 
so that the relation (\ref{ineq}) becomes an equality, as expected for
the pure-state case. This shows that the dynamics for the selected pulse
parameters satisfies the pulse selectivity conditions required for a
complementarity experiment.

A disadvantage of such a realization is the long overall duration of
the pulse sequence. A way to reduce this duration 
is to use a simple pulse-shaping technique and to replace the
long pulse with two short ones, separated by 
$\tau=\pi\hbar/\Delta $. With the FWHM of each of these
pulses of $0.1$ ps and the total duration of the pulse sequence
$t_{\mathrm{S}}=1.05$ ps a very good agreement with the ideal case
is achieved with the total duration reduced by an order of magnitude.

\section{Effect of environment-induced dephasing}
\label{sec:decoh}

Another important factor leading to reduction
of visibility are decoherence processes that take place between
the pulses. Such processes build up correlations between the confined
carriers and their environment so that the environment has some
\textit{which path} knowledge on the carrier state and the visibility
is decreased, even though no useful knowledge on the system state is
available, turning Eq.~(\ref{ineq}) into inequality.  In this Section,
the pulse parameters established above will be used to simulate the
system evolution in the presence of dephasing resulting from coupling
with the environment. Here we choose a simple, Markovian model of
dephasing that was successfully used to explain observed properties of
interface fluctuation QDs \cite{guenther02,li03}.

Dephasing may be understood by treating the environment
as a third party that continuously extracts information on both
subsystems during the experimental sequence. The fringe
contrast is decreased not only because some information has been
extracted by the
occupation measurement but also due to the portion of information
transfered to the environment. On the other hand, the dephasing of the
second subsystem amounts to sharing the information between the latter
and the environment. This does not affect the fringe visibility but
does affect the availability of information stored in the second
system \cite{sprinzak00}, 
so that experimentally accessible information that may be
used to predict the outcome of a measurement on the first subsystem is
now lower than the \textit{a priori} amount of $\sin(\alpha/2)$.

\begin{figure}[tb]
\begin{center}
\unitlength 1mm
\begin{picture}(85,65)(0,5)
\put(0,5){\resizebox{85mm}{!}{\includegraphics{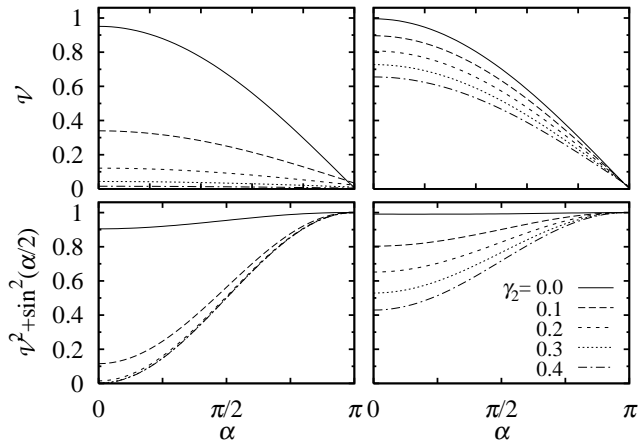}}}
\end{picture}
\end{center}
\caption{\label{fig:complem2}Upper: the exciton occupation fringe
visibility as a function of the \textit{a priori} amount of extracted
knowledge on the exciton occupation for 
$T_{1}=T_{2}/2=100$ ps and additional pure dephasing rates as
shown (in ps$^{-1}$). Lower: the distinguishability-visibility relation under
conditions as above. Left panels show the results for a long QP pulse
and right ones for a short double pulse.}
\end{figure}

The effects of dephasing are 
included within a Markovian model, assuming that the two
excitons are coupled to independent reservoirs, both characterized by
an occupation relaxation time $T_{1}=100$ ps and a dephasing time
$T_{2}=(\gamma_{1}+\gamma_{2})^{-1}$, where $\gamma_{1}=1/(2T_{1})$
and $\gamma_{2}$ accounts for pure
dephasing effects resulting, e.g., from fluctuating electrostatic
environment. The  
relevant Master equation \cite{breuer02} reads
\begin{equation}\label{master}
\dot{\rho}=-i[H,\rho]+D[\rho]
\end{equation}
with the Hamiltonian (\ref{ham0}) and the 
dissipator $D[\rho]=D_{1}[\rho]+D_{2}[\rho]$ consisting of
two parts. The first one accounts for the radiative decay of the
individual excitons,
\begin{displaymath}
    D_{1}[\rho]=\gamma_{1}\sum_{i=1}^{2}\left[
        \Sigma_{-}^{(i)}\rho\Sigma_{+}^{(i)}
        -\frac{1}{2}\{\Sigma_{+}^{(i)}\Sigma_{-}^{(i)},\rho
        \}_{+}\right],
\end{displaymath}
where $\Sigma_{\pm,z}^{(1)}=\sigma_{\pm,z}\otimes \mathbb{I}$,
$\Sigma_{\pm,z}^{(2)}=\mathbb{I}\otimes \sigma_{\pm,z}$.
The second contribution describes the additional pure dephasing
\begin{displaymath}
    D_{2}[\rho]=\gamma_{2}\sum_{i=1}^{2}\left[
        \Sigma_{z}^{(i)}\rho\Sigma_{z}^{(i)}
        -\frac{1}{2}\{\Sigma_{z}^{(i)}\Sigma_{z}^{(i)},\rho
        \}_{+}\right].
\end{displaymath}
The evolution equation (\ref{master})
is solved numerically for a range of pure
dephasing rates $\gamma_{2}$. Based on the results of the simulations,
visibility of interference fringes is calculated and  plotted in
Fig.~\ref{fig:complem2} both for long pulse and shaped
pulse case (left and right, respectively). Upper panels show the
visibility as a function of the QP pulse area (related to the \textit{a
priori} distinguishability of states). 
For long pulses, the
decrease of fringe visibility in the absence of exciton detection, due
only to dephasing, 
is noticeable even without the additional dephasing and becomes
dramatic as the latter is increased to the experimentally known values of
$\gamma_{2}\sim 0.1$ ps$^{-1}$. If a double short pulse is used to
reduce the 
sequence duration, dephasing is much weaker and the effect of
occupation relaxation is unnoticeable. As is clear from the lower
panels in Fig.~\ref{fig:complem2}, dephasing turns Eq.~(\ref{ineq})
into an inequality, except for $\alpha=\pi$, where a projective measurement
is performed on the first exciton and the fringe contrast is reduced
to zero anyway. Nonetheless, even for relatively strong dephasing, the
reduction of visibility due to which path information remains clearly
visible. 

\section{Conclusion}

It has been shown that the fundamental complementarity
between the wave-like properties of a quantum system (interference
effects) and its particle-like characteristics (the presence or absence of an
exciton, treated as an indivisible entity) may be quantitatively tested
in a time-domain interference experiment on
semiconductor quantum dots. The feasibility of such an experiment has
been confirmed by numerical simulation of system dynamics with
realistic parameters and under typical environmental dephasing, as
known from experiments.

Although an experiment studied theoretically in this paper demands
high level of coherent control over the quantum state of a confined
exciton, recent progress in the experimental studies of QDs 
\cite{krenner05}
makes such experiments feasible with the existing experimental
techniques in a number of systems. In particular, 
high degree of control over carrier states
in various kinds of QD structures has been demonstrated
\cite{zrenner02,stievater01,borri02a,unold04}, including phase control
necessary for interference experiments in non-perturbative regime
\cite{kamada01,htoon02} as well as coherent manipulations on a
biexciton system in a single QD or two coupled QDs \cite{li03,unold05}. 

The phenomena underlying the quantum complementarity demonstration
discussed in this paper are similar to those predicted in the case of
spontaneous bi-exciton decay via the two nearly-degenerate
single-exciton states \cite{hohenester03,economou05}, 
where information transfer due to entanglement 
with an emitted photon destroys coherence of the remaining exciton
state. In contrast to these works, the demonstration of quantum
complementarity requires that the correlations be induced in a
controlled way by exciting the system with a particular pulse
sequence. Hence, in this case correlations between excitons appear due
to external optical driving, while photon emission serves only as an
indication of the exciton presence.

The time-domain manifestation of quantum complementarity discussed here 
not only broadens
the class of experiments in which fundamental aspects of the quantum
world may be tested but also has the advantage of being independent
of the position-momentum (Heisenberg's) uncertainty that has been 
historically tied
to the space-domain discussions of complementarity \cite{wooters79}. 
In fact, it is independent of any uncertainty principles whatsoever.
Indeed, the only two quantities which are measured 
in the time-domain interference experiment are the occupations
of two different excitons. These quantities refer to \textit{different
subsystems} and, therefore, are obviously \textit{commuting} and
simultaneously measurable. Although the quantum probe exciton is
created in a way that correlates it with the presence of the other exciton
(S), this cannot be interpreted as a (projective) measurement on S, 
since the QP exciton is definitely a quantum (microscopic) system and
not a classical, macroscopic measurement device. Thus, the
experimental procedure described in this paper demonstrates quantum
complementarity in its pure form, involving only the notion of
\textit{information} on the system state and independent of any
uncertainty relations between non-commuting observables.

\section*{Acknowledgement}

The author is grateful to the Alexander von Humboldt Foundation
for support.
This work was supported by the Polish MNI under Grant No.
PBZ-MIN-008/P03/2003. 


\end{document}